\documentclass[prd,showpcs,amsmath,amssymb,showkeys,superscriptaddress]{revtex4-1}
\usepackage{color}
\usepackage{epsfig}

\newcommand{\e}{e}
\usepackage{graphicx}
\usepackage{dcolumn}
\usepackage{bm}

\begin{document}

\title{Construction of multiple spherical branes cosmological scenario
\\}

\author{I. C. Jardim}
\email{jardim@fisica.ufc.br}
\affiliation{Departamento de F\'{\i}sica, Universidade Federal do Cear\'{a},
Caixa Postal 6030, Campus do Pici, 60455-760, Fortaleza, Cear\'{a}, Brazil.}
\author{R. R. Landim}
\affiliation{Departamento de F\'{\i}sica, Universidade Federal do Cear\'{a},
Caixa Postal 6030, Campus do Pici, 60455-760, Fortaleza, Cear\'{a}, Brazil.}
\author{G. Alencar}
\affiliation{Universidade Estadual do Cear\'a, Faculdade de Educa\c c\~ao, Ci\^encias e Letras do Sert\~ao Central-R. 
Epit\'acio Pessoa, 2554, 63.900-000  Quixad\'{a}, Cear\'{a},  Brazil.}
\author{R. N. Costa Filho}
\affiliation{Departamento de F\'{\i}sica, Universidade Federal do Cear\'{a},
Caixa Postal 6030, Campus do Pici, 60455-760, Fortaleza, Cear\'{a}, Brazil.}

\date{\today}
           
\begin{abstract}
The general solution of Einstein's gravity equation in $D$ dimensions for an anisotropic
and spherically symmetric matter distribution is calculated in a bulk with position
dependent cosmological constant. Results for $n$ concentric $(D-2)-$branes with arbitrary
mass, radius, and pressure with different cosmological constant between branes are found.
It is shown how the different cosmological constants contribute to the
effective mass of each brane. It is also shown that the equation of state for each
brane influences the dynamics of branes, which can be divided into eras according to the
dominant matter. This scenario can be used to model the universe in the $D=5$ case, which
may presents a phenomenology richer than the current models. The evolution law of the
branes is studied, and the anisotropic pressure that removes divergences is found. The 
Randall-Sundrum metric in an outside the region in the flat branes limit is also derived.
\end{abstract}

\pacs{04.50.-h, 04.20.Jb, 04.50.Gh} 

\maketitle

\newpage
\section{Introduction}
The general model for the cosmos is based on the description of the universe as a perfect
fluid that admits a global cosmic time. This scenario has a space-time with constant curvature given by the Friedmann-Robertson-Walker metric, where  its dynamics is
determined by a cosmological scale factor that depends on the fluid state equation.
Despite the success of that model in the description of the primordial nucleosintesys and
the cosmic microwave background, it has failures that led to the emergence of new
models. Among the major flaws of the current model are the problem of dark energy showing
the accelerated expansion in the currently observed universe, and the dark matter which is
the divergence between the rotation of the halo of some galaxies and the amount of matter
contained in them according to gravitational dynamics \cite{weinberg:cosmology}.

Cosmological models with extra dimensions appeared first in Kaluza-Klein models with extra
dimensions and later in Randall-Sundrum scenarios \cite{Randall:1999vf, Randall:1999ee}.
These models describe the observed universe as a brane universe in a hyper-dimensional
space-time. Despite the fact that the Friedmann-Robertson-Walker metric did not determine
the geometry of the observed universe. The majority of studies focused on plane geometry.
Because of its  simplicity, this geometry is not able to change the dynamics of the
universe and thus cannot solve the problem of dark matter or the initial singularity.

Although the first studies to describe the universe as a spherical shell back to the 80's
\cite{Rubakov:1983bb,Visser:1985qm,Squires:1985aq}, the spherical brane-universe has
shown very rich phenomenology  in the past decade \cite{Gogberashvili:1998iu,Boyarsky:2004bu}.
Besides being compatible with the observational data
\cite{Tonry:2003zg,Luminet:2003dx,Overduin:1998pn}, the models provide an explanation for; the galaxy
isotropic runaway (isotropic expansion), the existence of a 
preferred frame, and a cosmic time. They show how the introduction of different
cosmological constants in each region of the bulk can change the dynamics of the
cosmological scale factor so as to make it compatible with the observed dynamics
\cite{Knop:2003iy,Riess:2004nr} without the introduction of dark energy
\cite{Gogberashvili:2005wy}. Similar to other models with extra dimensions, the spherical
shell models open the possible to obtain an energy scale in order to solve the problem of
hierarchy \cite{Gogberashvili:1998vx} and can be used as a basis for systems with varying
speed of light in the observed universe \cite{Gogberashvili:2006dz}.

The introduction of other branes and different cosmological constants can modify the
overall dynamics of the observed universe. Local density fluctuations of density can change the
local dynamics such as galactic dynamics (since the field of other branes interacts
gravitationally with the matter of the brane-universe) without  
dark matter. Herein this piece of work we extend and generalize the scenario of the world as one expanding shell \cite{Gogberashvili:1998iu} to multiple concentric
spherical $(D-2)$-branes in a $D$ dimensional space-time. For this, we solve the Einstein's
equation in $D$ dimensions to $n$ $(D-2)-$branes with different masses in a space with
different cosmological constants between the branes. A previous study considered a continuous  distributions of matter. However, only one cosmological constant was used \cite{Das:2001md}. We solve the  $D-$dimensional case, but for a cosmological model we limited ourselves to the case $D=5$, since the observed universe has only three spatial dimensions.
 
This work is organized as follows: In the second section we review the Einstein's equations in $D$ dimensions with a cosmological constant for spherically symmetric matter 
distribution. In the third section we solve this set of equations for $n$ shells with different cosmological constants $\Lambda$ between them. In Sec. 4, the
energy-momentum tensor conservation law is used to determine the possible anisotropic pressure
which removes the divergences in brane evolution equation. In the fifth section we particularize the solution found to take the flat brane limit in order to obtain the
Randall-Sundrum metric in the exterior region. In the last section we discuss the conclusions
and possible consequences.

\section{Static and Spherically Symmetric Space-time in $D$ Dimensions}
To learn about the gravitational effect of a distribution of matter we must determine the geometry of space-time. For this we need to know the $D(D+1)/2$ independent components of the 
metric solving the Einstein's equation. However, it is possible to use the symmetry of the problem to reduce
these components to just two, given by the invariant line element\cite{Gogberashvili:1998iu},
\begin{equation}
ds^{2} = -A(r,t)dt^{2} +B(r,t)dr^{2} +r^{2}d\Omega^{2}_{D-2}
\end{equation}
where $\Omega_{D-2}$ is the element of solid angle in $D$ dimensions, formed by $D-2$ angular variables.

Therefore we are left only with two functions, $A(r,t)$ and $B(r,t)$, to be determined by the Einstein's equation in $D$ dimensions
\begin{equation}\label{einD}
 R_{\mu}^{\nu} -\frac{1}{2}R\delta_{\mu}^{\nu} +\Lambda\delta_{\mu}^{\nu} = \kappa_{D}T_{\mu}^{\nu},
\end{equation}
where $\Lambda$ is the cosmological constant, which depends on $r$ and possibly  on $t$. Also $\kappa_{D}$ is the gravitational coupling constant in $D$ dimensions.
Due to the symmetries of the problem we only have four non null independent components of the Einstein's equation (\ref{einD}), which are
\begin{small}
\begin{eqnarray}
\kappa_{D}T_{0}^{0} &=&-\frac{D-2}{2r^{2}}\left[(D-3)\left(1 -B^{-1}\right) +\frac{rB'}{B^{2}}\right] +\Lambda \label{ein00}, \\
\kappa_{D}T_{1}^{1} &=&  -\frac{D-2}{2r^{2}}\left[(D-3)\left(1 -B^{-1}\right) -\frac{rA'}{AB}\right] +\Lambda \label{ein11},\\
\kappa_{D}T^{1}_{0} &=& \frac{D-2}{2r}\frac{\dot{B}}{B^{2}}, \label{ein10}\\
\kappa_{D}T_{2}^{2} &=& \frac{1}{4A}\left[\frac{\dot{A}\dot{B}}{AB} +\frac{\dot{B}^{2}}{B^{2}} -\frac{2\ddot{B}}{B}\right] +\frac{(D-3)(D-4)}{2Br^{2}} - \nonumber
\\&&-\frac{2(D-3)(D-4)}{r^{2}}  +\frac{(D-3)}{2Br}\left(\frac{A'}{A} -\frac{B'}{B}\right) + \nonumber
\\&&  +\frac{1}{4B}\left[\frac{2A''}{A} -\frac{A'^{2}}{A^{2}} -\frac{A'B'}{AB}\right]  +\Lambda \label{ein22},
\end{eqnarray}
\end{small}
where the prime means derivation with respect to $r$ and the dot is the derivative with respect to $t$.

We can see that if we know $T^{0}_{0}$, $T^{1}_{1}$ and $\Lambda$ we can, from (\ref{ein00}) and (\ref{ein11}), completely determine the solutions with two boundary conditions.
This comes from the fact that we have two first order differential equations. In this case the remaining equations determine the flow of energy $T^{1}_{0} $, and the tangential 
stresses $ T^{2}_{2}$. To find the exact solution we need to specify the form of matter $T^{\mu}_{\nu}$ which we use. 

\section{General Solution for Thin Spherical Branes }
The cosmological scenario we shall consider consists of $n$ concentric spherical delta type $(D-2)$-branes in a $D$ dimensional space with different cosmological constant
between them. As said in the introduction this is a generalization of \cite{Gogberashvili:1998iu}. For this we fix the energy-momentum tensor
and the cosmological constant to the form
\begin{equation}\label{fixT}
 T_{0}^{0}(r,t) = -\sum_{i=1}^{n}\rho_{i}\delta(r-R_{i}), \quad
 T_{1}^{1} = \sum_{i=1}^{n}P_{i}\delta(r-R_{i}),  
\end{equation}
and
\begin{small}
\begin{equation}
\Lambda(r,t) = \frac{(D-1)(D-2)}{2}\sum_{i=0}^{n}\lambda_{i}\left[\theta(r-R_{i}) -\theta(r -R_{i+1})\right],
\end{equation}
\end{small}
where the dependence on $t$ should be solely due to the branes radii ($ R_{i} = R_{i}(t) $). The $\theta$ function is defined in such way that is 1 when
the argument vanish, it's made to ensure that above expresion cover all space, including $r=0$ point.

The cosmological constant can be understood as a special fluid, so we can think that the difference between the cosmological constant is because each brane contains a fluid with different 
density. Fixing $T^{0}_{0}$ and $\Lambda$ we can find $B(r,t)$ using equation (\ref{ein00}) in the form
\begin{equation}
\kappa_{D}T^{0}_{0} = -\frac{D-2}{2r^{D-2}}\left[r^{D-3}\left(1 -B^{-1}\right)\right]' +\Lambda \label{ein002}
\end{equation}
and according to the above equation, $B$ has a first order discontinuity in $R_{i}$ because $T^{0}_{0}$ has a second order one and $\Lambda$ has first
order discontinuity only. 
Where we consider that in the region $R_{i} \leq r <R_{i +1}$, $B(r,t) = B_{i}(r)$, since in this region $B$ does not depend on $t$. The above-mentioned time dependence occurs in the region where this solution is valid. 

The region between the branes has no matter. Therefore, the equation (\ref{ein10}) assures us that the solution is static in this region. This information is contained in the Birkhoff's theorem. 
Integrating (\ref{ein002}) from $ R_{j} - \epsilon $ to $R_{j} + \epsilon$ and taking the limit $ \epsilon \to 0$, we obtain the discontinuity in the point $ r = R_ {j} $
\begin{equation}\label{descont}
 B_{j}^{-1}(R_{j}) -B_{j-1}^{-1}(R_{j}) = -\frac{2\kappa_{D}}{D-2}\rho_{j}R_{j}
\end{equation}
The limit $\epsilon \to 0$ eliminate the $\Lambda$ term because its divergence is first order only. Integrating (\ref{ein002}) from $R_{j} +\epsilon$ to
$r< R_{j+1}$, where $B$ is continuous, and taking the limit $\epsilon \to 0$ we obtain 
\begin{eqnarray*}
B_{j}^{-1}(r) &=&  1- \left(\frac{R_{j}}{r}\right)^{D-3}\left[1 -\lambda_{j}R_{j}^{2} -B_{j}^{-1}(R_{j})\right] -\lambda_{j}r^{2}
\\&=& 1- \left(\frac{R_{j}}{r}\right)^{D-3}\left[1 -\lambda_{j}R_{j}^{2} +\frac{2\kappa_{D}}{D-2}\rho_{j}R_{j} \right] + \\&& +\left(\frac{R_{j}}{r}\right)^{D-3}B_{j-1}^{-1}(R_{j})-\lambda_{j}r^{2}.
\end{eqnarray*}
By recurrence we find that
\begin{eqnarray*}
B_{j}^{-1}(r) &=&  1- \frac{1}{r^{D-3}}\sum_{i=1}^{j}\left[\frac{2\kappa_{D}}{D-2}\rho_{i}R_{i}^{D-2} -\Delta\lambda_{i}R^{D-1}_{i}\right] +
\\&&+ \left(\frac{R_{1}}{r}\right)^{D-3}\left(1-\lambda_{0}R_{1}^{2} -B_{0}^{-1}(R_{1})\right) -\lambda_{j}r^{2} ,
\end{eqnarray*}
where $\Delta\lambda_{i} =\lambda_{i} -\lambda_{i-1}$. Considering that inside all branes the solution is a de Sitter vacuum, i.e.,
$
B_{0}(r) =\left( 1-\lambda_{0}r^{2}\right)^{-1},
$
we get
\begin{eqnarray}
B_{j}^{-1}(r) &=& 1- \frac{1}{r^{D-3}}\sum_{i=1}^{j}\left[\frac{2\kappa_{D}}{D-2}\rho_{i}R_{i}^{D-2} -\Delta\lambda_{i}R^{D-1}_{i}\right] -\nonumber
\\&& -\lambda_{j}r^{2}.\label{Bj}
\end{eqnarray}
The above solution is valid only in the region  $R_{j}\leq r<R_{j+1}$, but we can write the solution valid in any region in the form
\begin{equation}
 B^{-1}(r,t) =  1- \frac{2G_{D}M(r,t)}{r^{D-3}} -r^{2}\lambda(r,t) \label{B} 
\end{equation}
where $M(r,t)$ and $\lambda(r,t)$ are defined by
\begin{small}
\begin{eqnarray*}
M(r,t) &\equiv & \sum_{i=0}^{n}\left[\frac{\kappa_{D}}{(D-2)G_{D}}\rho_{i}R_{i}^{D-2} -\frac{\Delta\lambda_{i}}{2G_{D}}R^{D-1}_{i}\right]\theta(r-R_{i}),
\\ \lambda(r,t) &\equiv& \sum_{i=0}^{n}\lambda_{i}[\theta(r -R_{i}) - \theta(r -R_{i+1})],
\end{eqnarray*}
\end{small}
and the time dependence is implicit in $R_{i}$. It is important to note that $M(r,t)$ is not positive defined, in order to enable a repulsive gravitational situation. Using the above definition in (\ref{ein11}) we find the equation which governs $A$
\begin{equation}
 \frac{A'}{A} = \frac{2\kappa_{D}}{D-2}BrT^{1}_{1} +2B\left[(D-3)G_{D}\frac{M(r,t)}{r^{D-2}} -r\lambda(r,t) \right]. \nonumber
\end{equation}
Taking the way $T^{1}_{1}$ was fixed at (\ref{fixT}), $A$ has a second order discontinuity. Now, using the same procedure to find $B$ we can show that
\begin{equation}
 A_{j}(r) = B_{j}^{-1}(r)A_{0}(R_{1})B_{0}(R_{1})\prod_{i=1}^{j}\frac{B_{i}(R_{i})}{B_{i-1}(R_{i})}e^{\pi_{i}}.\nonumber
\end{equation}
where
\begin{equation*}
\pi_{i} \equiv \frac{2\kappa_{D}}{D-2}R_{i}B_{i}(R_{i})P_{i}.
\end{equation*}
The asymptotic behavior of $B(r,t)$ is $\lim_{r\to\infty} B(r) = \left[ 1-\lambda(r)r^{2}\right]^{-1}$,
which is the generalization of the de Sitter vacuum to a cosmological constant that is position dependent. Likewise we expect that $ A(r) $ behaves asymptotically as the vacuum, i.e.,
$\lim_{r\to\infty} A(r) =  1-\lambda(r)r^{2}$, so that we can use this to fix the multiplicative constants appearing in the temporal solution and write
\begin{equation}
 A_{j}(r) = B_{j}^{-1}(r)\prod_{i=j+1}^{n}\frac{B_{i-1}(R_{i})}{B_{i}(R_{i})}e^{-\pi_{i}}.\nonumber
\end{equation}
In the same way we did for $B$, we can rewrite the above solution in order to be valid in all space as
\begin{eqnarray}
A(r,t) &=& B^{-1}(r,t)\prod_{i=1}^{n}\e^{-\pi_{i}\theta(R_{i} -r)}\times \nonumber
\\&& \times\left[1+\left(\frac{B_{i-1}(R_{i})}{B_{i}(R_{i})}-1\right)\theta(R_{i}-r)\right],\label{A}
\end{eqnarray}
where $B_{j}$ is defined by (\ref{Bj}).

The solutions (\ref{A}) and (\ref{B}) are generalizations of the Kottler solution \cite{Kottler-Ann.Phys.361} in $D$ dimensions with position dependent $\Lambda $.
In the case where $\lambda$ is constant, these solutions agree with those found by Das \cite{Das:2001md}. However, these solutions only make sense if the branes are not in a time-like region. In order to avoid a singularity in the solutions, we impose that the radius of the brane relates to the masses so that they are beyond their respective generalized Kottler's radii, i.e.
\begin{eqnarray}
-\lambda(R_{i}) R_{i}^{D-1} +R_{i}^{D-3} -2GM(R_{i}) >0. 
\end{eqnarray}
The above solutions  perfectly agree with the Birkhoff's theorem, and despite a constant, it is the Schwarzschild solutions with a cosmological constant (Kottler Solution). The temporal dependence of the solutions is in $R_{i}$, so that in each region the solution is static. 
The multiplicative constants in the temporal part of the solutions indicates the gravitational redshift, even within the shells. Mathematically this makes the 
solution continuous in all regions.

The time dependence on $B$ is given exclusively by $R_{i}$. Therefore in a dynamical case ($R_{i} = R_{i}(t)$) we can obtain the energy flow from (\ref{ein10}).
It's easy to show that
\begin{equation}
T^{1}_{0} = -\sum_{i=1}^{n}\rho_{i}V_{i}\delta(r -R_{i}),\nonumber
\end{equation}
where $V_{i} \equiv \dot{R}_{i}$.
The tangential stresses can be obtained from (\ref{ein22}), but it is easier to compute from the energy-momentum tensor conservation law. 

\section{Energy-Momentum Tensor Conservation Law}
The Einstein's field equation relates the energy-momentum tensor and the metric tensor. But due the symmetry only two components of energy-momentum tensor are necessary to determine
the metric. So the other terms of energy-momentum tensor are determined by Einstein's equation or by a conservation law. Taking the covariant derivative in the equation (\ref{einD}) we obtain the $D$ dimensional conservation law.
 \begin{eqnarray}
T_{\mu;\nu}^{\nu} = \frac{\Lambda_{,\mu}}{\kappa_{D}},
\end{eqnarray}
the above equation states that the energy and momentum are not conserved inside the brane because the extra dimensional pressure given by the difference between cosmological constants. This difference can be used to model the dark energy, which makes the universe expand. But in our case no strange matter in needed inside the brane like the usual dark matter models.
In terms of independent components the above conservation law can be written as
\begin{eqnarray}
\frac{\dot{\Lambda}}{\kappa_{D}} &=&\dot{T}^{0}_{0} +T^{1\prime}_{0} +\frac{\dot{B}}{2B}\left[T^{0}_{0} -T^{1}_{1}\right]+ \nonumber
\\&&+\frac{T^{1}_{0}}{2}\left[\frac{A'}{A} +\frac{B'}{B} +\frac{2(D-2)}{r}\right]
\\\frac{\Lambda'}{\kappa_{D}} &=&\dot{T}^{0}_{1} +T^{1\prime}_{1} +\left[\frac{A'}{2A} +\frac{(D-2)}{r}\right]T^{1}_{1}+ \nonumber
\\&& +\frac{T^{0}_{1}}{2}\left[\frac{\dot{A}}{A} +\frac{\dot{B}}{B}\right] -\left[\frac{A'}{2A}T^{0}_{0} +\frac{(D-2)}{r}T^{2}_{2}\right].
\end{eqnarray}
The first equation is trivially satisfied if we use the known solution (\ref{A}) and (\ref{B}) and energy-momentum tensor components. The second equation gives us the propagation speed of the brane as a function of the 
tangential stress, the masses, and the cosmological constant in the same way as in (\ref{ein22}). Taking
\begin{equation}
T^{2}_{2} = \sum_{i=1}^{n}T_{i}\delta(r-R_{i})\nonumber 
\end{equation}
we can integrate the unsolved component of the conservation law from $R_{i}-\epsilon$ to  $R_{i}+\epsilon$ to obtain   
\begin{eqnarray*}
\frac{\Delta\Lambda_{i}}{\kappa_{D}} &=& \frac{B_{i}(R_{i})}{A_{i}(R_{i})}[\dot{\rho}_{i}V_{i} +\rho_{i}\dot{V}_{i}] +\frac{D-2}{R_{i}}\left[P_{i} -T_{i}\right]+ 
\\&&+\left[\left.\left(P_{i} +\rho_{i}\right)\frac{A'}{2A} +\rho_{i}V_{i}^{2}\left(\frac{B'}{A} -\frac{BA'}{A^{2}}\right)-\right.\right. 
\\&&-\left.\left.\rho_{i}V_{i}\left(\frac{B\dot{A}}{2A^{2}} -\frac{3\dot{B}}{2A}\right)\right]\right|_{r=R_{i}}
\end{eqnarray*}
as the functions $A$ and $B$ have second order divergences in $r = R_{i}$ the last term in above expression have the same divergence. Analyzing separately the divergent terms
\begin{widetext}
\begin{eqnarray*}
\mbox{div} &=&  \overbrace{\frac{\kappa_{D}}{D-2}B_{i}(R_{i})R_{i}\left(P_{i} +\rho_{i}\right)\left[P_{i}-\frac{B_{i}(R_{i})}{A_{i}(R_{i})}\rho_{i}V_{i}^{2} \right]\left.\delta(r-R_{i})\right|_{r=R_{i}}}^{\mbox{real divergence}} +
\\&&+B_{i}(R_{i})\left[P_{i} +\rho_{i} -4\rho_{i}V_{i}^{2}\frac{B_{i}(R_{i})}{A_{i}(R_{i})}\right]\left[(D-3)\frac{G_{D}M(R_{i})}{R_{i}^{D-2}} -R_{i}\lambda_{i}\right] -\rho_{i}V_{i}\frac{B_{i}(R_{i})}{2A_{i}(R_{i})}\left[K_{0}(R_{i}) -\sum_{j=i}^{n}\dot{\pi}_{j}\right]
\end{eqnarray*} 
\end{widetext}
where
\begin{equation}
K_{0} = \sum_{j=1}^{n}\left[\frac{\dot{B}_{j-1}(R_{j})}{B_{j-1}(R_{j})} -\frac{\dot{B}_{j}(R_{j})}{B_{j}(R_{j})} \right]\theta(R_{j}-r).\nonumber
\end{equation}
To avoid a real divergence we need to fix
\begin{equation}
P_{i} = -\rho_{i} \;\;\;\;\;\mbox{or}\;\;\;\;\;  P_{i} =\frac{B_{i}(R_{i})}{A_{i}(R_{i})}\rho_{i}V_{i}^{2}.\nonumber
\end{equation}
The first case indicates a cosmological constant state equation type, this equation is the only state that is independent of motion, i.e. the properties of a fluid with this state equation is independent of its movement.
Therefore, it was already expected that the divergences found in the dynamic case can be removed. The second case relates the normal pressure with the brane velocity. That indicates an increase in the pressure if the velocity increases to keep the spherical shape of the brane. This relationship ensures that $P$ could vanishes in the static case, as in the Randall-Sundrum scenario. Assuming a linear state equation relating the tangential stresses and the energy density,
$
T_{i} = \gamma_{i}\rho_{i},
$
and defining, for the $i$-th brane, the time
\begin{equation}
dt_{i} = \sqrt{A_{i}(R_{i})/B_{i}(R_{i})}dt \nonumber
\end{equation}
the brane evolution is given by
\begin{eqnarray}
\rho_{i}\frac{dU_{i}}{dt_{i}}&=& \frac{\Delta\Lambda_{i}}{\kappa_{D}}\left(1- U_{i}^{2}\right) -\frac{D-2}{R_{i}}\left[P_{i} -\rho_{i}\left(\gamma_{i} +U_{i}^{2}\right)\right] - \nonumber
\\&&-B_{i}(R_{i})\left[P_{i} +\rho_{i} -2\rho_{i}U_{i}^{2}\right]\times \nonumber
\\&&\times\left[(D-3)\frac{G_{D}M(R_{i})}{R_{i}^{D-2}} -R_{i}\lambda_{i}\right],\label{evolution}
\end{eqnarray} 
where
$
U_{i} \equiv \frac{d R_{i}}{dt_{i}} .
$
This indicates a different dynamic for each cosmological eras driven by a different state equation , i.e., by the related tangential pressures.

\section{The Randall-Sundrum Flat Brane Limit}
In the previous sections we found the general solution to $n$ spherical branes in a $D$-dimensional space-time with different cosmological constant between them. To find a scenario
similar to Randall-Sundrum we need to fix $D=5$, $n=1$ and $\lambda_{0}=\lambda_{1}$. In this case the exterior solution is
\begin{equation}
ds^{2} = -f(r)dt^{2}  +f^{-1}(r)dr^{2} +r^{2}d\Omega_{3}^{2},\nonumber 
\end{equation}
where
\[ f(r)= \left(1 -\frac{2G_{5}M}{r^{2}} -\lambda r^{2}\right).\]
In order to obtain the Randall-Sundrum metric we define 
\begin{equation}
dz \equiv \left(1 -\frac{2G_{5}M}{r^{2}} -\lambda r^{2}\right)^{-1/2}dr\nonumber
\end{equation}
or, fixing that $z$ vanishes when $r =R$, 
\begin{equation}
z = \frac{1}{2k}\ln\left[\frac{2k\left(k^{2}r^{4} +r^{2} -2G_{5}M\right)^{1/2} +2k^{2}r^{2} +1}{2k\left(k^{2}R^{4} +R^{2} -2G_{5}M\right)^{1/2} +2k^{2}R^{2} +1}\right],\nonumber
\end{equation}
where $\lambda =-k^{2}$ to avoid the de Sitter horizon. To obtain the flat brane limit we will consider that $R$ and consequently $r$ tends to infinity.
In this limit the dominant term, regarding that $M$ grows as $R^{3}$, is
\begin{equation}
z = \frac{1}{2k}\ln\left[\frac{r^{2}}{R^{2}}\right]\nonumber
\end{equation}
writing $r$ as function of $z$, we obtain the line element 
\begin{equation}
ds^{2} = -k^{2}\e^{2kz}R^{2}dt^{2}  +dz^{2} +\e^{2kz}R^{2}d\Omega_{3}^{2}.\nonumber
\end{equation}
Putting the constants into coordinates we obtain the Randall-Sundrum metric
\begin{equation}
ds^{2} = \e^{2kz}\eta_{\mu\nu}dx^{\mu}dx^{\nu}  +dz^{2}.\nonumber
\end{equation} 
The exponential in the warp factor is positive because the bulk is anti de Sitter, instead the original RS scenario.    

\section{Conclusions and Perspectives}
In this work we built a scenario of multiple concentric membranes through the solution of 
Einstein's equation in $D$ dimensions with different cosmological constant in each region. The results
 we found may serve as a basis for more specific scenarios, through the fixation of radii, masses on 
each brane, and bulk cosmological constant. In a dynamical case, the solutions we found can be used to 
model the universe for $D=5$.  The model used here is more accurate than the previous ones, and as a consequence 
we have a multiplicative constant that appears in the temporal solution, which is the redshift measured by 
observers in the region inside the brane. 

Through the momentum-energy tensor conservation law we obtained two possible anisotropic pressures that remove 
the singularities in the branes dynamics. These two possible pressures give us two possible fixations leading 
to more freedom in the construction of a cosmological model that can better fits the observed data. The tangential 
pressure and the difference between the cosmological constants are responsible for the evolution of each brane
 according to Eq. (\ref{evolution}). This tangential pressure is found from a state equation that is determined by 
the dominant matter in each cosmological era.   We show that the difference between the cosmological constants 
modifies the effective mass of the matter distribution, and can be fixed in a way that the observed universe 
expansion rate is independent of the dark energy. Finally, we were able to  arrive at the Randall-Sundrum metric 
to a anti-de Sitter bulk from calculating the external solution in the limit of plane branes. This metric was 
first introduced in the literature as an ansatz\cite{Randall:1999ee}. Now, it is derived from the Kottler anti-de 
Sitter solution. The main extension of  the model developed here is the phenomenological study of cosmology 
generated by solving a dynamic Universe Brane equation.

\begin{acknowledgements}
We would like to thank the financial support provided by Funda\c c\~ao Cearense de Apoio ao Desenvolvimento Cient\'\i fico e Tecnol\'ogico (FUNCAP), the Conselho Nacional de 
Desenvolvimento Cient\'\i fico e Tecnol\'ogico (CNPq), and FUNCAP/CNPq/PRONEX.
\end{acknowledgements}


\begin{thebibliography}{5}

\bibitem{weinberg:cosmology}
  S.~Weinberg,
  {\it Cosmology},
  Oxford University Press Inc., New York (2008)

\bibitem{Randall:1999vf}
  L.~Randall and R.~Sundrum,
  Phys.\ Rev.\ Lett.\  {\bf 83}, 4690 (1999)
  [arXiv:hep-th/9906064].

\bibitem{Randall:1999ee}
  L.~Randall and R.~Sundrum,
  ``A large mass hierarchy from a small extra dimension,''
  Phys.\ Rev.\ Lett.\  {\bf 83}, 3370 (1999)
  [arXiv:hep-ph/9905221].

\bibitem{Rubakov:1983bb}
  V.~A.~Rubakov and M.~E.~Shaposhnikov,
  Phys.\ Lett.\  B {\bf 125}, 136 (1983).

\bibitem{Visser:1985qm}
  M.~Visser,
  Phys.\ Lett.\  B {\bf 159}, 22 (1985)
  [arXiv:hep-th/9910093].

\bibitem{Squires:1985aq}
  E.~J.~Squires,
  Phys.\ Lett.\  B {\bf 167}, 286 (1986).

\bibitem{Gogberashvili:1998iu}
  M.~Gogberashvili,
  Europhys.\ Lett.\  {\bf 49}, 396 (2000)
  [arXiv:hep-ph/9812365].

\bibitem{Boyarsky:2004bu}
  A.~Boyarsky, A.~Neronov and I.~Tkachev,
  Phys.\ Rev.\ Lett.\  {\bf 95}, 091301 (2005)
  [arXiv:gr-qc/0411144].

\bibitem{Tonry:2003zg}
  J.~L.~Tonry {\it et al.}  [Supernova Search Team Collaboration],
  Astrophys.\ J.\  {\bf 594}, 1 (2003)
  [arXiv:astro-ph/0305008].

\bibitem{Luminet:2003dx}
  J.~P.~Luminet, J.~Weeks, A.~Riazuelo, R.~Lehoucq and J.~P.~Uzan,
  Nature {\bf 425}, 593 (2003)
  [arXiv:astro-ph/0310253].

\bibitem{Overduin:1998pn}
  J.~M.~Overduin and P.~S.~Wesson,
  Phys.\ Rept.\  {\bf 283}, 303 (1997)
  [arXiv:gr-qc/9805018].

\bibitem{Knop:2003iy}
  R.~A.~Knop {\it et al.}  [Supernova Cosmology Project Collaboration],
  Astrophys.\ J.\  {\bf 598}, 102 (2003)
  [arXiv:astro-ph/0309368].

\bibitem{Riess:2004nr}
  A.~G.~Riess {\it et al.}  [Supernova Search Team Collaboration],
  Astrophys.\ J.\  {\bf 607}, 665 (2004)
  [arXiv:astro-ph/0402512].

\bibitem{Gogberashvili:2005wy}
  M.~Gogberashvili,
  Phys.\ Lett.\  B {\bf 636}, 147 (2006)
  [arXiv:gr-qc/0511039].

\bibitem{Gogberashvili:1998vx}
  M.~Gogberashvili,
  Int.\ J.\ Mod.\ Phys.\  D {\bf 11}, 1635 (2002)
  [arXiv:hep-ph/9812296].

\bibitem{Gogberashvili:2006dz}
  M.~Gogberashvili,
  Europhys.\ Lett.\  {\bf 77}, 20004 (2007)
  [arXiv:hep-th/0603235].


\bibitem{Das:2001md}
  A.~Das and A.~DeBenedictis,
  Prog.\ Theor.\ Phys.\  {\bf 108}, 119 (2002)
  [arXiv:gr-qc/0110083].

\bibitem{Kottler-Ann.Phys.361}
  F.~Kottler,
  Ann.\ der Phys.\ {\bf 361}, 14 (1918).

\bibitem{Birkhoff}
  G.~D.~Birkhoff,
  {\it Relativity and Modern Physics},
  Harvard University Press: Boston (1923).

\end{thebibliography}
\end{document}